# Report of the Kavli IAU Workshop on Global Coordination: Future Space-Based Ultraviolet-Optical-Infrared Telescopes

Debra Elmegreen, Ewine van Dishoeck, David Spergel, and Roger Davies

## I. INTRODUCTION

International efforts play a key role in driving all areas of astrophysics, and involve access to data and facilities and joint partnerships on large-scale instruments, observatories, and missions. Global cooperation and collaboration are increasingly important as the costs and preparation to build forefront facilities and missions escalate. International strategic planning is essential to explore how partnerships can be built and joint projects developed that otherwise could not be afforded, and to maximize the scientific return from these efforts.

For these reasons, the International Astronomical Union (IAU) established in 2016 the Working Group on Global Coordination of Ground and Space Astrophysics, reporting directly to the Executive Committee (EC), following an IAU GA 2015 Focus Meeting by the precursor Division B Working Group on Large Scale Facilities. Its co-chairs are David Spergel and Roger Davies; the WG members are listed at:
https://www.iau.org/science/scientific_bodies/working_groups/278/members/
and the EC liaisons are Ewine van Dishoeck and Debra Elmegreen. A key part of the WG activities is to organize meetings to foster international planning. The IAU facilitates discussions, but *does not endorse* any particular mission or facility more or less than another. The aim of the WG is to have one Focus Meeting at every General Assembly covering a broad range of issues, and one focused workshop in between GAs.

An IAU Workshop on "Global Coordination: Future Space-Based Ultraviolet-Optical-Infrared Telescopes," generously sponsored by the Kavli Foundation, was the first such meeting. The 3-day meeting took place at Kasteel Oud Poelgeest in Leiden, the Netherlands on July 17-19, 2017 with 40 invited participants from 17 countries (Australia, Brazil, Canada, China, Denmark, France, Germany, India, Israel, Italy, Japan, Netherlands, Russia, South Africa, Spain, United Kingdom, United States). Participants from universities, observatories, research organizations, and space agencies included science community leaders and experts on various aspects of relevant science and technology. This report is a brief summary of the workshop. For more details, the meeting agenda and slides from presentations, along with background documentation, are available at: https://www.strw.leidenuniv.nl/KavliIAU2017

The workshop sessions were centered on 4 themes: (1) Setting the stage - astronomy in the 2030s, (2) Science drivers - why do we need a large UVOIR mission?, (3) Visions for large missions - what capabilities do we need?, and (4) How large missions fit into long-term space mission plans; these are summarized below. The ultimate aim of this workshop was to fertilize discussions about the development of ambitious joint space missions, since long-term prioritization processes around the globe are ramping up again. This report may help inform White Papers for upcoming decadal surveys and roadmaps.

This workshop also provides input to the Focus Meeting on "Global Coordination of International Astrophysics and Heliophysics Activities from Space and Ground," to be held at the 2018 General Assembly.

## II. Motivation: the big picture

One of the most fundamental questions that humankind has posed over the centuries is whether we are alone in the Universe. A prime driver for a large UVOIR space telescope is to search for biomarkers from habitable Earth-like planets around solar-type stars. As Nobel Laureate Riccardo Giacconi noted in 1997, 21$^{st}$ century astronomers should be uniquely positioned to be able to tackle observationally "the evolution of the Universe in order to relate causally the physical condition during the Big Bang to the development of RNA and DNA." The science behind this grand endeavor encompasses a broad range of astrophysics, along with atmospheric physics as well as geological and biological topics. If a large UVOIR mission ultimately succeeds in determining with high confidence that features in an exoplanet's atmosphere are indicative of life and cannot be explained by other non-biological mechanisms, the mission would be one kind of success. Conversely, if a complete search of habitable planets around nearby solar-type stars reveals no such biomarkers, that would provide an important constraint limiting the likelihood of remotely detectable life, which would also be a critical achievement. For major facilities, being able to achieve a solid null result is just as important as actual detection and should be part of the design strategy. Current estimates for placing high confidence limits on such observations indicate that a space-based telescope must have a diameter larger than 10 meters (>12-meter preferred) to reach a sufficient sample size. Such a telescope would also enable access to studies of a wide range of other critical astrophysical questions beyond detecting habitable Earths, some of which are noted below.

Beyond having compelling science drivers, it is crucial for any facility comparable to the James Webb Space Telescope (JWST) and the Hubble Space Telescope (HST) to garner widespread community support for such a mission. Moreover, it is imperative to plan for international contributions, coordinate ancillary efforts in technology developments, theoretical advancements, and simulations, and consider political, budgetary, and complexity factors that go into a multinational effort.

## III. Setting the stage - astronomy in the 2030s

Multi-wavelength observations, surveys with follow-up observations, and time domain studies all lend themselves to coordinated approaches to astronomical research. The astronomy landscape in the 2030s will include extremely large ground-based optical telescopes, sub-millimeter and radio arrays, and high-energy facilities and gravitational wave observatories on the ground and in space (see presentations). The expense and effort involved in developing a large UVOIR space mission only makes sense if it can accomplish unique and complementary science in the context of other current and future facilities.

## IV. SCIENCE DRIVERS - WHY DO WE NEED A LARGE UVOIR MISSION?
### A. EXOPLANETS

A large UVOIR mission would be unparalleled in the study of possible habitable planets. It would allow direct detection of Earth-like planets and spectral characterization of their atmospheres, something beyond the capability of the JWST or the planned WFIRST coronagraph. Such spatially-separated exoplanetary spectra are fundamentally different from transit or eclipse spectroscopy, where the planetary signal is only a tiny fraction of the total signal and which is generally limited to planets close to their parent star. Direct imaging requires a large diameter telescope, as the inner working angle of a coronagraph will typically be several $\lambda/D$. ELTs on the ground will have diameters up to 39 meters and thus small Inner Working Angles (IWA) down to 10 milliarcsec. With an expected planet/star contrast ratio of $10^{-8}$, these ELTs could likely search for biosignatures in the atmospheres of habitable Earth-like planets around M stars. With our current understanding of Adaptive Optics technologies, the stabilities required, and the brightness and relative paucity of G-type stars compared to M-type stars, the ELTs may be challenged from the ground to do the same for solar-type G stars. From space, contrast ratios of $10^{-10}$ would be more readily achievable, which would enable a 10-meter-class space telescope to characterize potential exoEarths around G stars. However, due to the larger IWA, such a space telescope could not target the habitable zones of many M stars. Thus, ground and space would be highly complementary in probing the full range of G-M stars that might harbor habitable exoplanets. Figure 1 shows the planet/star contrast for FGK stars versus inner working angle or apparent separation and the limits achievable for different telescopes.

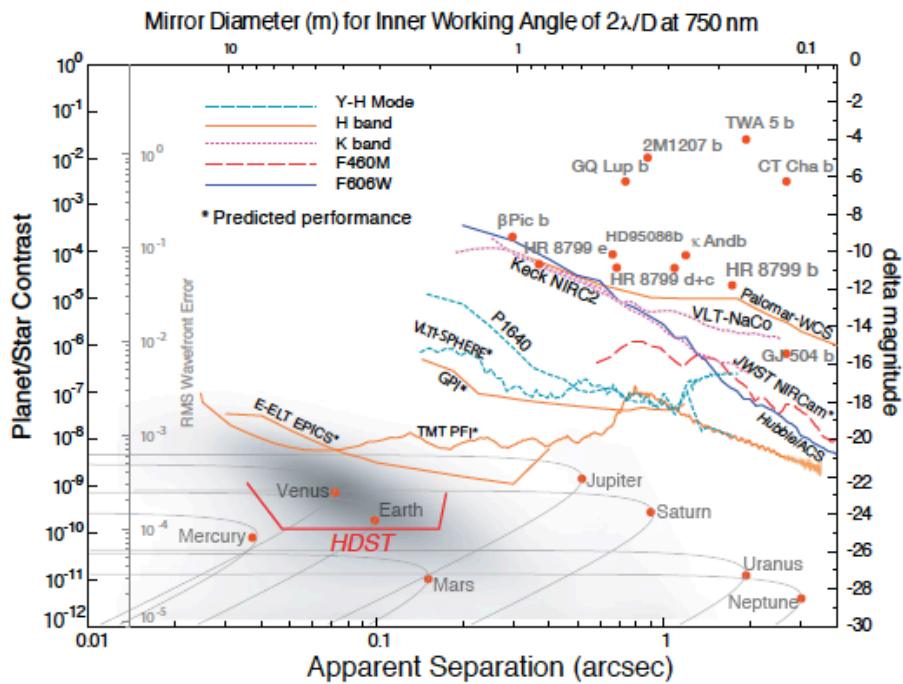

*Figure 1: Direct imaging contrast capabilities of current and future instrumentation, from AURA's 2015 report "From Cosmic Birth to Living Earths: The Future of UVOIR Space Astronomy." HDST (High Definition Space Telescope) represents a large UVOIR telescope.*

Prior to the launch of a large UVOIR telescope, there would need to be an input catalogue of candidate exoplanets in habitable zones already determined from other surveys. There are many current ground and space efforts to detect exoplanets from direct imaging, radial velocity, transit, astrometry, and microlensing observations that will help form the survey sample for a large UVOIR mission by the time it would be launched.

The number of exoplanets whose atmospheres would need to be characterized in order to have a statistically significant result depends on the value of $\eta_{Earth}$, the fraction of GKM stars with Earth-like planets in their habitable zones. According to current estimates, a space telescope with a diameter of 12 meters or larger could reach a statistically significant sample size for characterizing diversity. With a complete search, it would be possible to report with high confidence the detection of molecular absorption features in the 0.6-2.5 μm range indicative of life in an exoplanet atmosphere, such as a combination of $O_2$, $O_3$, $CH_4$, $H_2O$, $CO_2$ (that could not be explained by known false positives), or that no planets within the search distance exhibit these signs of life. A null result from such a census would be as important (though perhaps not as exciting) as a positive result, since it would constrain the frequency of remotely detectable life.

A search for life-supporting atmospheres would also enable studies of the range of atmospheric processes and planet evolution histories, the physics and chemistry of the nearest exoplanets, chemical build-up in atmospheres, global climate and weather, and possible reflection signatures through direct imaging and spectroscopy, maybe including large exomoons too.

**B. COSMIC ORIGINS**

There are compelling science drivers beyond exoplanet atmospheres that would benefit from a large UVOIR mission. Some key science drivers for having UV coverage (preferably down to 120 nm) include:
- Origin and evolution of the elements
- Origin and evolution of the cosmic web
- Evolution of the baryon cycle

The material from which rocky planets are made originates as the product of stellar evolution, so tracing the origin of the elements ties directly into our place in the Universe. Such an effort includes far extragalactic and first light observations to help inform chemical and ionization history of the Universe by studying the onset, evolution, and truncation of different elements across cosmic time. The entire zoo of explosive and ejection sources that give rise to heavy elements with cosmic time could be catalogued using stars and gas signatures in Damped Lyman-alpha (DLA) objects and Lyman limit systems. Determining a DLA mass spectrum would provide insight into protogalaxy formation evolution. This history is related to the cosmic web, including understanding of the baryon mass and angular momentum.

The baryon cycle can be studied through Intergalactic medium (IGM)/circumgalactic medium (CGM) mapping of H I, C IV, N V, O VI, and other UV emission and absorption tracers, which helps understand galaxy evolution as baryons are accreted onto and expelled from galaxies, and explores processes such as accretion and quenching through feedback via winds expelling gas from galaxies. HST can study the CGM only with a limited number of sightlines, and neither JWST nor WFIRST has UV sensitivity; high spatial resolution multi-object emission mapping of the CGM would be truly unique. The local stellar record would help astronomers understand the dominant physical processes at $z > 1$. Star formation histories in different environments could be explored through resolved populations and from voids to massive filaments and clusters, pushing to the Virgo Cluster and beyond.

### C. ADDITIONAL AND COMPLEMENTARY SCIENCE

A wide range of science could be done across the UVOIR wavelengths with a large space telescope. UV spectroscopy in particular is a realm that would be unique to space. In addition, broad wavelength coverage out to the near IR would allow observations of the formation and evolution of the first objects in the epoch of reionization, from seed black holes to Population III stellar populations, in unprecedented physical detail. A 12-meter telescope would resolve a region 100 pc in size out to the limits of the observable Universe. Following up on observations with JWST, a large UVOIR telescope would help in characterizing earliest galaxy and black hole formation. JWST will transform our physical understanding of reionization and early galaxies, and will provide the first gas-phase metallicities, which are related to early AGN activity and gas outflows. After JWST's 5-10 year predicted lifetime and surveys planned from WFIRST and Euclid, a large UVOIR telescope will be critical for the next phase of probing early galaxies and detecting Population III supernovae and star clusters.

A large UVOIR telescope would be important for imaging protoplanetary disk structures down to a few AU. This capability would aid searches for young embedded planets in disks: according to exoplanet statistics every disk must be forming planets, but so far they have been very difficult to detect directly. Also, the origin and evolution of other, non-habitable planets could be explored by studying their atmospheric composition and linking it to their birth location. The composition of planetesimals could be probed in a unique way through spectroscopy of white dwarfs with infalling planetary debris. Within the Solar System, a large UVOIR telescope could tackle cosmogonic and planetary dynamics problems though spectroscopy and imaging of trans-Neptunian objects (TNO), near-Earth objects (NEO), and deep external Solar System objects, which would have a synergy with LSST observations. In the solar vicinity <100 pc, the initial mass function (IMF) could be probed down to even lower brown dwarf masses, making the connection with any free-floating giant planet population.

Massive stars are fundamental to many aspects of astrophysics, from feedback in galaxies to gravitational wave precursors. Metallicity is a key factor in their structure and evolution, and only a large UVOIR space telescope could undertake a systematic spectroscopic study of the nearest 'first galaxy' analogs. UV is critical for determining precise ages. A large UVOIR telescope could enable a systematic spectroscopic survey of massive stars out to distances of 10s of Mpc in different environments and physical conditions. Through spectroscopy of high mass binaries, a large UVOIR telescope would be able to study the local counterparts of the early Universe explosions, including superluminous supernovae that may be responsible for the first observable photons. Studying white dwarfs would help in understanding the ages and roles they play in Type Ia SNe and help test fundamental physics.

There is a broad range of complementary science for a large UVOIR telescope working in tandem with ground and space telescopes. Important parallel programs (especially for transient science) include wide field, high spatial resolution and sensitivity studies of precursors and aftermaths of merging black holes and binary neutron stars detected with observatories such as advanced-LIGO and Virgo, gamma ray bursts (GRBs) and X-ray bursts (XRBs) observed with Fermi and Athena, fast radio bursts (FRBs) observed in the radio with UTMOST, LOFAR, and SKA, optical transients observed with LSST, Gaia, JWST, WFIRST, and Euclid, and testing of fundamental constants (e.g., examining possible variations of the fine structure constant through high spectral resolution and sensitivity from 120-400 nm).

The history of science has shown that whenever there is a quantum leap in telescope apertures, sensitivity, and resolution limits, perhaps the greatest science to be done is in the realm of discovery. Invariably, large facilities with multiple capabilities will be doing observations not even imagined originally. With the quest for exoEarths and the corollary observations across the astronomical realm driving the mission, the unexplored unknowns that await discovery will likely yield even greater returns on the investment.

## V. VISIONS FOR LARGE MISSIONS - WHAT CAPABILITIES DO WE NEED?

The primary wavelength range for spectral characterization of exoplanet atmospheres is in the near IR, out to 2.5 µm. There are also good biomarker features at longer wavelengths in the mid-IR at 5-15 µm, but such observations require an actively cooled IR system and would best be done in a separate IR-optimized mission.

Characterizing spectral features in the atmosphere of an Earth-like planet at ~10 parsecs will require a 10+ meter telescope. This large aperture is needed both for its ability to take a spectrum of a $30^{th}$ magnitude Earth-like planet and for its small Inner Working Angle. The space environment is likely essential for the high contrast ratios needed to characterize planets around FGK stars. Technical capabilities needed include space-qualified active/adaptive optics, coronagraphs (perhaps including starshades), and low to moderate spectral resolutions (resolving power R=100-2000). Many of these technologies are being tested on WFIRST.

Key desiderata for ICM/CGM, stellar, and galaxy studies include imaging and high resolution IFU spectroscopy (to R~50,000) down to 120 nm, with a preferred stretch to 100 nm. CGM and stellar abundance diagnostics increase dramatically going to the EUV. A 10-meter or larger telescope (12- to 15-meter preferred) could target large numbers of star fields and be used to create redshift survey slices to probe the CGM. Enhanced high performance reflective coatings for mirrors, gratings, and detectors at UV wavelengths would be needed.

While this workshop did not aim for consensus and specific recommendations, the following basic guidelines summarize the needs presented for a large UVOIR space mission with a prime driver of exoEarth detection and a host of additional science and synergistic observations: a deployable 10-meter or larger segmented mirror with a broad wavelength coverage (e.g., 120 nm to 2.5 μm), diffraction-limited imaging in the optical and IR, multi-object (a few 100) IFUs with high spatial and spectral resolution spectroscopy in the UV, and moderate spectral resolution spectroscopy in the IR (R>2000).

In terms of technology development, the following challenges were identified: enhanced UV coatings; high contrast ($<10^{-10}$) coronography (possibly a starshade) with an <0.1" Inner Working Angle using a segmented telescope; light path control stability to a new level of precision. In addition, high quantum efficiency, low noise detectors (>32 AB mag sensitivity); signal processing algorithms, an end-to-end simulator, and efficient operation would all be critical to develop.

**VI. HOW LARGE MISSIONS FIT INTO LONG-TERM SPACE MISSION PLANS**

A large UVOIR mission is a critical complement to the suite of missions underway in many realms, which taken together provide broad wavelength coverage from space. Besides the impending launch of JWST, among other missions, NASA has WFIRST in the planning stages, ESA has approved JUICE (a large mission) for Jupiter and icy moon studies, Euclid (an O/IR medium mission) for dark energy studies, and PLATO for exoplanet transits with longer periods than TESS. Athena, a large X-ray observatory, has been selected and adopted, and the gravitational wave mission LISA has been selected for a future large mission. UV missions have been proposed to ESA but not selected. JAXA has recently launched ARASE to study radiation in geospace, and has a large number of other missions under consideration, including the ESA-led SPICA cryogenic infrared mission. It has approved the X-Ray Astronomy Recovery Mission (XARM; with NASA and ESA). Among other projects, Russia is in the process of building a 1.7-meter telescope (World Space Observatory-Ultraviolet) in collaboration with Spain. India has launched its AstroSat mission, which has simultaneous UV, optical, low and high energy X-ray detection. India is also developing an X-ray polarimeter satellite (XPoSat), along with several other ground and space missions and collaborations. China is planning a 2-meter UV telescope (Chinese Space Station Optical Survey, CSSOS) for high resolution near-UV imaging and slitless spectroscopy in the optical, in the same orbit as the Chinese Space Station for serviceability.

With ESA's cost cap of 1 billion Euros for large missions and similar cost caps for other agencies, a large UVOIR mission would need to be led by the U.S. but would have to be an international effort. Large missions historically have been in the planning stages for 2 decades or more, so the time is right to consider such a mission. A large UVOIR space telescope as a general-purpose observatory with multiple compelling science drivers (as in the cases of HST and JWST) would help engender broad community support. Such support is necessary to impact political and budgetary considerations.

Data on the costs of previous large missions support the conclusion that the cost of a mission does not scale directly with size or mass. The cost projection of HST (including repairs and servicing missions) scaled to a 6.5-meter telescope indicates that JWST is cheaper. Similarly, a UVOIR observatory is expected to be cheaper than a JWST scaled to a 10-meter or 12-meter telescope, due to advances in technologies and associated reductions in cost (and a UVOIR mission would not need cryogenics). The lessons learned from previous missions suggest that improvements in technology and new heavy launch vehicles (e.g., NASA's SLS) could help limit the cost of a large UVOIR telescope considerably.

As with HST, astronauts might play a significant role in the construction or operation of this large telescope. A serviceable observatory might make it more attractive too, to the extent that it could be adapted to rapidly changing science and technologies through replacement or refurbishment of instruments. The great success of HST stemmed from the ability to repair and improve it over time. Serviceability could engage the human endeavors side of NASA and other space agencies. In terms of launch mechanisms, a deployable, segmented mirror along the lines of JWST, in a suitably large launch shroud, could make sense.

Global coordination of an ambitious large UVOIR space telescope requires that each country or organization participate at an appropriate level and reasonable way. Its implementation by NASA will only have a chance if it is the top-ranked space mission in the next U.S. decadal survey. Other countries, including those with emerging economies, could contribute to technological development and/or instrument manufacturing. It is likely that many of the needed high-risk technologies could align well with the plans of private companies and industry.

The success of a large international effort requires effective management to deal with a large number of international stakeholders, both in terms of operations and data management. Realistically estimating and controlling the costs are critical. To this end, it is vital to achieve suitable technical readiness levels to predict and adhere to the estimated costs. Limiting the number of instruments and observing modes could be necessary; the mission should be ambitious but realistic. Successful science returns will also depend on prior appropriate models and simulations, including large-scale hydrodynamic simulations, laboratory astrophysics, and continued development of the theoretical framework needed for instrument design, optimizing science programs, and interpreting data, for which input from all countries will be useful.

## VII. RECOMMENDATIONS

A successful push for a large UVOIR mission demands that the compelling science cases be further developed and advocated. Such groundwork is underway through a variety of workshops and by a large number of science teams with international members.

Several studies in the U.S. over the last quarter century have outlined the case and advocated for a large UVOIR telescope. NASA is currently enabling four studies of future large-scale telescopes, including one for UVOIR capabilities ("LUVOIR"). The working groups for the Science and Technology Definition Team for LUVOIR and the other three mission concepts are open to new team members from any nation. There are occasional NASA related science conferences and webinars on particular aspects of such a mission; an exoplanet technology meeting is planned for April 2018.

Participants at this workshop recognized the utility and necessity of bringing groups with particular expertise together. Important take-aways include the need to pursue multiple options for different technologies and instruments and to continue discussions and advocacy for a large UVOIR mission.

Therefore, we recommend and urge that astronomers worldwide intensify their activities to explore the possibilities for science with a large UVOIR mission and, if such a mission is deemed compelling, to determine the roles that they might fill in instrumentation and technology definition and development, construction, launch, data analysis, and complementary science. Such groundwork is critical for the successful selection and development of a large UVOIR mission, and now is the time to take action.

## ACKNOWLEDGMENTS

We gratefully acknowledge the Kavli Foundation for its generosity in sponsoring this workshop. We thank the participants for their efforts in presentations and discussions during the workshop and in preparing this report, all of which contributed to the success of the meeting and provided the impetus to continue the conversations and work ahead.